\documentclass[hyper,a4paper]{JHEP}
\usepackage{amsmath,graphicx,epsfig,amssymb,multirow}

\allowdisplaybreaks

\preprint{arXiv:1601.xxxxx\\ \today}
\title{Anisotropic models are unitary: A rejuvenation of standard quantum cosmology }
\author{Sridip Pal$^{a,b,1}$,  Narayan Banerjee$^{b,2}$\\
$^a$Department of Physics, University of California, San Diego,
9500 Gilman Drive, La Jolla, CA 92093, USA\\
$^b$Department of Physical Sciences,
Indian Institute of Science Education and Research Kolkata,
Mohanpur, 741246, West Bengal, India\\
Email:~$^1$\email{srpal@ucsd.edu ; sridippaliiser@gmail.com}, $^2$\email{narayan@iiserkol.ac.in} }
\abstract{The present work proves that the folk-lore of the pathology of non-conservation of probability in quantum anisotropic models is wrong. It is shown in full generality that all operator ordering can lead to a Hamiltonian with a self-adjoint extension as long as it is constructed to be a symmetric operator, thereby making the problem of non-unitarity in context of anisotropic homogeneous model a ghost. Moreover, it is indicated that the self-adjoint extension is not unique and this non-uniqueness is suspected not to be a feature of Anisotropic model only, in the sense that there exists operator orderings such that Hamiltonian for an isotropic homogeneous cosmological model does not have unique self-adjoint extension, albeit for isotropic model, there is a special unique extension associated with quadratic form of Hamiltonian i.e {\it Friedrichs extension}. Details of calculations are carried out for a Bianchi III model.}
\begin{document}
\section{Introduction}
Any scheme of quantization of a cosmological model suffers from a severe problem as the anisotropic models were found to be non-unitary and hence failed to conserve probability\cite{nelson1}. Although the observed universe is isotropic, but this pathology definitely pulls the steam out of the quantization scheme for being non-trustworthy. Certainly this apparent non-unitary evolution is one of a long list of conceptual problems in quantum cosmology. There are many comprehensive reviews in this connection\cite{nelson1, wilt, halli}. It is interesting to note that this non-unitarity is often apt to be invisible in the absence of a properly oriented scalar time parameter in the scheme of quantization\cite{lidsey, nelson2}. In a relativistic theory, time itself is a coordinate and fails to be the scalar parameter against which the evolution is studied. In fact the problem of the proper identification of time in quantum cosmology is a subject by itself and discussed in detail by many\cite{kuchar1, isham, rovelli, anderson}.\\

It has been shown that using Schutz formalism\cite{schutz1, schutz2} of presenting the fluid degrees of freedom in terms of thermodynamic potentials, the evolution of the fluid can give rise to a properly oriented time parameter. The method was developed long back by Lapchinskii and Rubakov\cite{rubakov}. Following this idea, it had been shown by Alvarenga {\it et al}\cite{alvarenga3} that Bianchi I models suffer from the non-unitarity. However, it was shown that for Bianchi V model, the alleged non-unitarity can possibly be ``alleviated'' in the sense that the probability becomes a constant at later stages of the evolution\cite{barun}. Very recently, it has been categorically shown with explicit examples that Bianchi I\cite{sridip1}, Bianchi V and IX\cite{sridip2} models as well as a Kantowoski-Sachs model\cite{sridip3} can all have unitary evolution for a proper choice of operator ordering. Even when an explicit solution is not found, the existence of a self-adjoint extension for the Hamiltonian can be ascertained\cite{sridip1, sridip2, sridip3}. Furthermore, this is acheieved not at the cost of the anisotropy itself\cite{sridip4}. With a scalar field distribution, Almeida {\it et al} discussed the possibility of a self-adjoint extension for the Hamiltonian\cite{almeida}. \\

The present work shows that the myth of a generic non-unitarity of quantum anisotropic homogeneous cosmological models is not correct at all. All ordering does in fact result in a Hamiltonian which has a self-adjoint extension so as to admit a unitary evolution. The work also points out that anisotropic models are still somewhat special with the non-existence of {\it Friedrichs extension} which has some distinct features as discussed later. This kind of extension is possible for isotropic models only. It deserves mention that except for a few cases, the Wheeler-de Witt equation for anisotropic homogeneous model is difficult to be solved analytically, sometimes the concerned differential equation may not be separable as well. Hence, it is not always possible to do an explicit self-adjoint extension and this is exactly where lies the importance of a proof of existence which shows , no matter how we choose the operator order, there exist self-adjoint extension(s) of Hamiltonian, governing the evolution of anisotropic homogeneous quantum cosmologial models. It deserves mention that there are operators no having self-adjoint extension, but they do not have any sensible interpretation as a Hamiltonian in quantum mechanics. Hence, if some quantisation scheme leads to Hamiltonian, not having a slef-adjoint extension, then that brings question to the validity of scheme itself. Thus we are establishing the fact that the standard prescription of quantizing a cosmological model cannot be ruled out as inconsistent. Since, it is not always possible to solve the differential equation and hence construct the self-adjoint extension explicitly, it is not possible to investigate the cosmological implications of different extensions except for some general remarks, which we will discuss later.\\

The rest of the paper is organised as follows. In section II, we have two subsectios , first of which deals with Bianchi-III model with a proof of unitarity. In the next subsection, the proof is generalised so as to be applicable to all anisotropic homogoneous model for all operator ordering.  In the light of this proof, the claims made earlier in the literature regarding non untarity of Bianchi-I and discrepancy between Many world interpretations and de-Broglie Bohm interpretation stand corrected and we list out them as implications of our work. This subsection also contains general remarks how existence of self-adjoint extension(s) can possibly affect the physical implications of the model. The paper ends with a discussion and concluding remarks regarding unitary of quantum anisotropic homogoneous models, (Non)uniqueness of self-adjoint extensions and pointing out excatly how it differs from isotropic one regarding existence of {\it Friedrichs} extention.\\

\section{Unitarity in Bianchi III \& Other Anisotropic Models in general}
\subsection{Bianchi-III}
We start with the standard Einstein-Hilbert action
\begin{equation}\label{1}
\mathcal{A}=\int_{M}d^{4}x \sqrt{-g}R +2 \int_{\partial M} \sqrt{h}h_{ab}K^{ab}+\int_{M} d^{4}x \sqrt{-g}P,
\end{equation}

where $R$ is the Ricci scalar, $K^{ab}$ is the extrinsic curvature, and $h^{ab}$ is the induced metric over the boundary $\partial M$  of the 4 dimensional space-time manifold $M$, $g$ is the determinant of the metric over manifold $M$ and $P$ is the pressure of the fluid. The units are so chosen that $16\pi G = 1$.\\

The Bianchi type III model is given by the metric 
\begin{equation}
\label{Misner}
ds^{2}=n^{2}dt^{2}-e^{2\sqrt{3}\beta_{+}}dr^{2}-e^{-2\sqrt{3}\left(\beta_{+}+\beta_{-}\right)}\left[d\theta^{2}+\sinh^{2}\left(\theta\right)d\phi^{2}\right].
\end{equation}
Here lapse function $n$ and $\beta_{+}$ and $\beta_{-}$ are functions of time $t$. The reason for writing the metric in this form\cite{bastos} instead of a cartesian form is to bring out the close similarity between Bianchi III and the Kantowski-Sachs cosmology. In the latter, the hyperbolic coefficient of $d\phi^{2}$ is replaced by a sinusoidal function. \\

Given the action \eqref{1}, the Hamiltonian for the gravity sector can be written as
\begin{equation}
\label{hamiltoniangr}
H_{g}=\frac{n}{24}e^{\sqrt{3}\left(\beta_{+}+2\beta_{-}\right)}\left[-p_{\beta_{-}}^{2}+p_{\beta_{+}}^{2}+48e^{-2\sqrt{3}\beta_{-}}\right].
\end{equation}
With the widely used Schutz formalism of writing the fluid parameters in terms of thermodynamic variables\cite{schutz1, schutz2}, and using that to define a time evolution, the Hamiltonian for the fluid sector can be written as: 
\begin{equation}
\label{hamiltonianfl}
H_{f}=ne^{\alpha\sqrt{3}\left(\beta_{+}+2\beta_{-}\right)}p_{T}.
\end{equation}
The net (super) Hamiltonian is given by $H=H_{g}+H_{f}$ and a variation with respect to $n$ yields the Hamiltonian constraint,
\begin{equation}
\label{constraint}
e^{\sqrt{3}\left(1-\alpha\right)\left(\beta_{+}+2\beta_{-}\right)}\left\{-p_{\beta_{-}}^{2}+p_{\beta_{+}}^{2}+48e^{-2\sqrt{3}\beta_{-}}\right\}+24p_{T}=0.
\end{equation}\\

{\bf Stiff Fluid: $\alpha=1$} \\
As an example, we choose a stiff fluid given by $\alpha=1$. On quantization, we have the Wheeler-DeWitt equation,
\begin{equation}
\label{wd}
\left\{\frac{\partial^{2}}{\partial\beta_{-}^{2}}-\frac{\partial^{2}}{\partial\beta_{+}^{2}}+48e^{-2\sqrt{3}\beta_{-}}\right\}\Psi =24\imath\frac{\partial\Psi}{\partial T},
\end{equation}\\
in units of ${\hbar}=1$. The canonical momenta $p_{i}$ and $p_{T}$ are replaced by $-\imath \frac{\partial}{\partial  \beta_{i}}$ and $-\imath \frac{\partial}{\partial T}$ respectively \cite{alvarenga3, barun, sridip1}. \\

By a separation of variables, $\Psi=\phi(\beta_{-})\psi(\beta_{+})e^{-\imath ET}$, equation \eqref{wd} can be cast into
\begin{eqnarray}
\label{wd1}
\left\{\frac{\partial^{2}}{\partial\beta_{-}^{2}}+3k_{+}^{2}+48e^{-2\sqrt{3}\beta_{-}}\right\}\phi =24E\phi ,\\
\label{wd2}
\left[\frac{\partial^{2}}{\partial\beta_{+}^{2}}+3k_{+}^{2}\right]\psi = 0.
\end{eqnarray}

Now, with the definition, $||\psi ||\equiv\int_{-\infty}^{\infty} d\beta_{+}\psi\psi^{*}$, \eqref{wd2} is self-adjoint and it can be shown that norm for the $\beta_{+}$ sector is time independent and finite by explicit construction of wavepacket along the line of \cite{sridip3}.\\

We recast the equation \eqref{wd1} in the standard self-adjoint form, using the variable $\chi\equiv e^{-\sqrt{3}\beta_{-}}$, 
\begin{equation}
\frac{d}{d\chi}\left(\chi\frac{d\phi}{d\chi}\right) +\left(16\chi -\frac{8E-k_{+}^{2}}{\chi} \right)\phi =0,
\end{equation}
with definition of inner product being given by $\langle \phi_{1}| \phi_{2} \rangle \equiv \int_0^{\infty} d\chi \ \chi \  \phi_{1}^{*}(\chi) \phi_2(\chi).$
Hence, the Hamiltonian for $\beta_-$ sector is self-adjoint as well, ensuring a unitary time evolution. \\

{\bf General perfect fluid: $\alpha\neq 1$}

Here also, we follow the trick adopted in the case of a Kantowoski-Sachs model\cite{sridip3}. We propose following operator ordering: 

\begin{eqnarray}
\label{wd3}
\nonumber[-e^{\frac{\sqrt{3}}{2}\left(1-\alpha\right)\left(\beta_{+}+4\beta_{-}\right)}\frac{\partial}{\partial \beta_{+}}e^{\frac{\sqrt{3}}{2}\left(1-\alpha\right)\beta_{+}}\frac{\partial}{\partial \beta_{+}}+e^{\sqrt{3}\left(1-\alpha\right)\left(\beta_{+}+\beta_{-}\right)}\frac{\partial}{\partial \beta_{-}}e^{\sqrt{3}\left(1-\alpha\right)\beta_{-}}\frac{\partial}{\partial \beta_{+}} \\ +48e^{-2\sqrt{3}\beta_{-}}e^{\sqrt{3}\left(1-\alpha\right)\left(\beta_{+}+2\beta_{-}\right)}]\Psi =24\imath\frac{\partial\Psi}{\partial T}.
\end{eqnarray}

We effect a transformation of variables as $\chi_{+}\equiv e^{-\frac{\sqrt{3}}{2}\left(1-\alpha\right)\beta_{+}}\ \& \ \chi_{-}\equiv e^{-\sqrt{3}\left(1-\alpha\right)\beta_{-}}$, and use separability ansatz $\Psi= \phi(\chi_{+},\chi_{-}) e^{-\imath E T}$ to obtain

\begin{equation}\label{wd5}
H_{g}\phi =-\frac{1}{\chi_{-}^{2}}\frac{\partial^{2}\phi}{\partial\chi_{+}^{2}}+\frac{1}{\chi_{+}^{2}}\frac{\partial^{2}\phi}{\partial\chi_{-}^{2}}+48\chi_{-}^{\frac{2\alpha}{1-\alpha}}\chi_{+}^{-2}\phi = 24E\phi.
\end{equation}
Now it is easy to see that one can use Neumann's theorem which states that \\

{\it
``A symmetric operator $\hat{A}$ defined on domain $\mathcal{D}$  has equal deficiency index, if there exists a norm preserving anti-unitary conjugation map $C:\mathcal{D}\rightarrow\mathcal{D}$ such that $[\hat{A},C]=0$, which, in turn, shows that $\hat{A}$ admits self-adjoint extension''.} \\

Here $H_{g}$ satisfies the conditions for this theorem to be applied with $C$ being the map which takes $\phi$ to $\phi^{*}$ and hence admits self-adjoint extension i.e unitary evolution. There is an easy way to understand the theorem, which we include here for clarity. The deficiency indices $n_{\pm}$ are defined to be dimension of eigenspaces ($\mathcal{N}_{\pm}$) of $H_{g}$ with eigenvalue $\pm\imath$. Now we see,
\begin{eqnarray}
H_{g}\Psi_{\pm} =\pm \imath\Psi_{\pm}\\
\Rightarrow CH_{g}\Psi_{\pm} = \mp \imath C\psi; \ Since\ C\ is\ antiunitary\\
\Rightarrow H_{g} (C\Psi_{\pm})=\mp \imath (C\Psi_{\pm}); \ Since\ [H_{g},C]=0\ 
\end{eqnarray}
Thus we have shown that for every $\Psi_{\pm}\in\mathcal{N}_{\pm}$, we have $C\Psi_{\pm}\in \mathcal{N}_{\mp}$ i.e bijection between two eigenspaces, which show $n_{+}=n_{-}$. For a detailed discussion on the theorem and its applicability, we refer to the standard text by Reed and Simon\cite{reed}. It deserves mention that the same analysis holds for the Kantowski-Sachs model as well, with a minor difference in one signature in the expression for $H_{g}$\cite{sridip3}. \\

The rationale behind choosing the operator ordering as in \eqref{wd3} can now be explained. With this ordering, we have the kinetic term $\frac{\partial^{2}\phi}{\partial \chi_{\pm}^{2}}$ multiplied with $\chi_{\mp}^{2}$. Hence, the condition for $H_{g}$ being symmetric is same as the condition for a standard Laplacian to be symmetric, since the derivative with respect to $\chi_+$ term is multiplied with $\chi_-$ and vice versa, i.e we have following condition, 
\begin{equation}
\label{assum}
\left[\phi\frac{\partial\phi^{*}}{\partial \chi_{\pm}}-\phi^{*}\frac{\partial\phi}{\partial \chi_{\pm}}\right]_{0}^{\infty}=0.
\end{equation}
Thus the ordering plays a role in making $H_{g}$ a symmetric operator. Once it is guaranteed to be a symmetric operator, the self-adjoint extension is obvious following Neumann's theorem. It deserves mention that although the particular operator ordering in \eqref{wd3} is a sufficient condition for making $H_{g}$ symmetric, yet it is not a necessary one and this brings in the question of non-uniqueness of operator ordering and a hope that we might possibly be able to generalise the proof for all operator ordering, and for all anisotropic homogeneous models.\\

\subsection{Generalisation}
The idea of generalisation stems from following important realizations. We note that we can always define a complex conjugation map $C:\mathcal{Hi}\rightarrow\mathcal{Hi}$ such that it is norm preserving, since whatever be the definition of norm, it involves $\phi\phi^{*}$, hence does not change under $C$. As long as $H_{g}$ is real,  $CH_{g}=H_{g}C$ is satisfied trivially. Hence, if we have a symmetric $H_{g}$, we can always have a self-adjoint extension of such operators. Now, it is trivial to realise that almost all operator ordering lead to a symmetric $H_{g}$ and thus this statement is not specific to Bianchi-III, this is true for other anisotropic models as well. \\

For example, the non-unitarity reported by Alvarenga {\it et al}\cite{alvarenga3} comes under a scrutiny in the light of this theorem. Even in their prescribed ordering, the operator should have self-adjoint extension, though the extension might be difficult to be realized in practice. They apparently showed that the deficiency indices of the Hamiltonian of Bianchi-I cosmological model are unequal, hence, we do not have a self-adjoint extension even if the Hamiltonian is Hermitian to start with. The Hamiltonian, used by them is written here for clarity,
\begin{equation}\label{al}
H_{g}\psi =e^{3(\alpha-1)\beta_{0}}\left(\frac{\partial^{2}}{\partial \beta^{2}_{0}}-\frac{\partial^{2}}{\partial\beta^{2}_{+}}-\frac{\partial^{2}}{\partial\beta^{2}_{-}}\right)\psi = 24\imath \frac{\partial \psi}{\partial T}.
\end{equation}
However, it is easy to check that the Hamiltonian \eqref{al} used by Alvarega {\it et al}\cite{alvarenga3} does satisfy all the requirements for Neumann's theorem to be applied. Hence, Hamiltonian does have equal deficiency indices, thereby admits a self-adjoint extension. In this case, the speculation is that if we do the self-adjoint extension, that would make boundary condition more stringent which will eventually cast out all the states with time dependent norm from the Hilbert space.\\

Here we redo the deficiency index calculation of $H_{g}$  given by \eqref{al}, we seek solutions to the following differential equations,
\begin{equation}
H_{g}\psi_{\pm} =\pm \imath \psi_{\pm}
\end{equation}
and the solutions are given by,
\begin{eqnarray}
\psi_{+}=\phi\left(\beta_{\pm}\right)\left[a_{1}H^{(1)}_{\nu}(e^{\imath\frac{\pi}{4}}\chi)+a_{2}H^{(2)}_{\nu}(e^{\imath\frac{\pi}{4}}\chi)\right],\\
\psi_{-}=\phi\left(\beta_{\pm}\right) \left[c_{1}K_{\nu}(e^{\imath\frac{\pi}{4}}\chi)+c_{2}I_{\nu}(e^{\imath\frac{\pi}{4}}\chi)\right],
\end{eqnarray}
where $\phi=e^{\imath\left(k_{+}\beta_{+}+k_{-}\beta_{-}\right)}$, $\chi=\frac{2}{3\left(1-\alpha\right)}e^{\frac{3}{2}\left(1-\alpha\right)\beta_{0}}$ and $\nu=\frac{\imath \sqrt{k_{+}^{2}+k_{-}^{2}}}{\frac{3}{2}\left(1-\alpha\right)}$.  From the asymptotic expansion, it can be verified that $I_{\nu}$ and $H^{(2)}_{\nu}$ blow up at infinity while $H^{(1)}_{\nu}$ and $K_{\nu}$ do not. Hence, the deficiency indices are given by $n_{+}=n_{-}=1$, and thus Neumann's theorem asserts that a self-adjoint extension is possible for the operator given by Alvarenga {\it et al}\cite{alvarenga3}.\\

Before venturing into a general proof, we list out things which stand corrected in light of our work in context of Bianchi-I model, and in fact, anisotropic models in general.
\begin{enumerate}
\item Unitarity guarantees that there is no discrepancy between Many World inpterpretation/ Copenhagen interpretation and de Broglie-Bohm interpretation. Alvarenga et. al. \cite{alvarenga3} argued that there is a  nonequivalence between the two interpretations owing to non-unitarity of the model, which, according to them, is due to hyperbolic structure of kinetic term in Wheeler-deWitt equation. It deserved mention that self-adjoint Hamiltonian leads to no discrepency between Many World inpterpretation/ Copenhagen interpretation and de Broglie-Bohm interpretation is a necessary and sufficient condition \cite{alvarenga3}. Hence, the conclusion regarding the discrepancy made by Alvarenga et. al. is now corrected in the light of the present work.

\item A related cosmological implication is regarding late time bahviour of Bianchi-I universe. Alvarenga et. al. found that the expectation value of the scale factor reveals an isotropic universe in late time when calculated in the spirit of Many World interpretation while bohmian trajectories revealed anisotropy is present even in late times. Now that there is no such discrepency, two interpretation should yield identical results. In \cite{sridip4}, we have been able to show that isotropy can be achieved at late times for Bianchi-I.
\end{enumerate} 

In general, for an anisotropic homogeneous model, the classical Hamiltonian can be written as
\begin{equation}
\sum_{i,j}A_{ij}\left(\beta_{k}\right)p_{i}p_{j}+V\left(\beta_{k}\right)=cp_{T}.
\end{equation}
We can always work in a co-ordinate system where $A_{ij}$ is diagonal i.e $A_{ij}=A_{i}\delta_{ij}$ (no summation implied). Although an example of Bianchi III is given here, and Bianchi I, V, IX and Kantowski-Sachs models were mentioned in \cite{sridip1,sridip2, sridip3}, this can explicitly be worked out for all the anisotropic homogeneous models, which in fact come either under Bianchi classification or is a Kantowoski Sachs Model. In the desired co-ordinate system, upon quantization, we have
\begin{equation}\label{add}
H_{g}\Psi = \left[-\sum_{i}F_{i}\left(\beta_{k}\right)\partial_{i}G_{i}\left(\beta_{k}\right)\partial_{i}+V\left(\beta_{k}\right)\right]\Psi = cE\Psi
\end{equation}
where $F_{i}$ and $G_{i}$ specifies the particular operator ordering such that $A_{i}=F_{i}G_{i}$ (no summation implied) and they are functions of the superspace co-ordinate $\beta_{k}$, where $k$ runs from $1$ to $p$, $p$ being the number of independent co-ordinates we require.  It is easy to see that $H_{g}$ is symmetric. The claim can be proved in following way:\\

Without loss of generality we can assume, $G_{i}$ is a function of $\beta_{i}$ alone since we can pull out the part depending on $\beta_{j\neq i}$, pass it to the left through $\partial_{i}$ and absorb in $F_{i}$. Now we effect a change of variable via
\begin{equation}
d\chi_{i}=G_{i}^{-1}d\beta_{i}
\end{equation}
and recast \eqref{add} in following form:
\begin{equation}
H_{g}\Psi = \left[-\sum_{i}\frac{F_{i}}{G_{i}}\frac{\partial^{2}}{\partial\chi_{i}^{2}}+V\left(\chi_{k}\right)\right]\Psi = cE\Psi
\end{equation}
Now $H_{g}$ is a symmteric operator with following defintion of norm,
\begin{equation}
||\Psi ||=\int \prod_{i} d\chi_{i}\ \left(\prod_{i}\frac{G_{i}}{F_{i}}\right) \Psi^{*}\Psi
\end{equation}

Now the antiunitary norm preserving map $C$ as defined earlier ($C\psi=\psi^{*}$) commutes with $H_{g}$ and thus Neumann's theorem goes through and $H_{g}$ admits self-adjoint extension(s), though it may not be unique. The choice of $F_i$ and $G_i$ is arbitrary as long as $A_{i}=F_{i}G_{i}$ Hence, for all operator ordering, we have shown that unitarity is guaranteed.\\

Now that we have shown the existence of self-adjoint extension for all anisotropic homogeneous models, the natural question is to ask how the different extensions affect the cosmology of the concerned model. However, specifcally commenting on how a particular extension effects a particular model requires solving Wheeler-deWitt equation, which can be tricky for many cases. For Bianchi-I, the extension can be done explicitly for a particular operator ordering\cite{sridip1,sridip4}. In case of Binachi-V, IX, Kantowoski Sachs (KS) model, it is possible to do it explicitly only for some choices of the perfect fluid, for a particular operator ordering\cite{sridip2,sridip3}. For KS and Bianchi-III, we see the differential equation is not even separable for $\alpha\neq1$ fluid. In spite of these technical difficulties, we can take cue from Bianchi-I studied in\cite{sridip4} to make following remarks:
\begin{enumerate}
\item Self-adjoint extension essentially controls the behaviour of wavefunction near singularity when sclae factor shrinks to $0$. It employs a boundary condition on the wavefunction.
\item The boundary condition(s) can result to a specific energy specturm of the Hamiltonian of the universe, can even set a ground state energy for the universe, which is what happens in Bianchi-I \cite{sridip4}.
\item In case of wavefunction representing a superposition of collapsing and expanding universe, the ratio of amplitude of birth of a new universe and collapse of a universe is dictated by self-adjoint extension so as to conserve probability and maintain unitarity.
\end{enumerate}

\section{Discussion and conclusion}
The work conclusively shows that for anisotropic models, all operator ordering can lead to a Hamiltonian, which admits self-adjoint extension. Hence, non-unitarity is not a problem at all. However, it is suspected that the extension is not unique in the context of an anisotropic model. All the models where we have been able to calculate the deficiency index e.g Bianchi-I in ordering scheme prescribed in \cite{sridip1} (henceforth called {\it NS} ordering), Bianchi-V, IX in NS \cite{sridip2} ordering, Bianchi-I in ordering used by Alvarenga {\it et al}\cite{alvarenga3} show that the extension is not unique and characterised by a $U(1)$ group. Hence, it is reasonable to ask \\

{\it Is the non-uniqueness of self-adjoint extension an exclusive feature of anisotropic model only or is there a possibility that similar situation can occur even in Isotropic models?} \\

It deserves mention that in \cite{sridip2}, isotropic models are claimed to suffer from nonunitairty upon particular ordering and that argument was based on that presented by Alvarenga {\it et al}\cite{alvarenga3}. Hence, the statement that isotropic model suffer from non unitarity for following ordering:
\begin{equation}\label{so}
e^{3\left(\alpha -1\right)\beta_{0}}\frac{1}{24}\frac{\partial^{2}\Psi}{\partial \beta_{0}^{2}}=\imath\frac{\partial\Psi}{\partial T}
\end{equation}
also stands wrong in the light of this theorem. It is interesting to note that although the ordering \eqref{so} admits self-adjoint extension, the extension is not unique. We can follow the same steps as we did with the Hamiltonian \eqref{al}, with $k=\sqrt{k_{+}^{2}+k_{-}^{2}}=0$, since we do not have separate $\beta_{\pm}$ sectors in isotropic models and arrive at $n_{\pm}=1$. This example corroborates to the general result that non unitarity can not be associated with bad choice of operator ordering, unitarity is preserved for all operator ordering. \\

Similarly, the following NS ordering \cite{sridip2}
\begin{equation}\label{soo}
\left[e^{\frac{3}{2}\left(\alpha -1\right)\beta_{0}}\frac{\partial}{\partial\beta_{0}}e^{\frac{3}{2}\left(\alpha -1\right)\beta_{0}}\frac{\partial}{\partial\beta_{0}}\right]\Psi =24 \imath\frac{\partial\Psi}{\partial T}
\end{equation}
can also be shown to have a one parameter $U(1)$ family of self-adjoint extensions. Hence, the answer to the question raised is no in general, we can have operator orderings, which can actually make the self-adjoint extension of isotropic model, non-unique. However, for isotropic models, $-H_{g}$ is a positive symmetric operator. This aids the use of {\it Friedrich's Theorem} \cite{reed}, which states that \\

{\it
Let $A$ be a positive symmetric operator  and let $q\left( \phi ,\psi\right) = \langle \phi , A\psi\rangle$ for $\phi,\psi \in D(A)$. Then $q$ is a closable quadratic form and its closure $\tilde{q}$ is the quadratic form of a unique self-adjoint operator $\tilde{A}$, such that lower bound of its spectrum is lower bound of $q$ and $\tilde{A}$ is the only self-adjoint extension of $A$, whose domain is contained in the form domain of $\tilde{q}$.} \\

It deserves mention that {\it Friedrich's extension} is unique in the sense that it has some nice relationship with the quadratic form associated with the operator as stated in the theorem. The mathematical aspect of the theorem is explained in the appendix for the sake of completeness as this theorem is not too frequently used in physics. The nicest feature about Friedrichs extension is that it is enough to characterize the boundary conditions to make sure $H_{g}$ is a symmetric operator. Friedrich's extension turns out to be a self-adjoint extension with the same specified boundary conditions. On the other hand, in the context of Anisotropic models, using Neumann's theorem makes sure that symmetric operator $H_{g}$ has self-adjoint extension but in practice the extension process involves modifying the boundary conditions (i.e making boundary conditions which make the operator symmetric, stronger), and this often involves introducing parameters by hand, if the extension is not unique. If we only want extension without such nice relationship with quadractic forms, then just being a positive symmetric operator does not guara ntee unique self-adjoint extension (for example, $H_{g}$ in \eqref{so} and \eqref{soo} does not have a unique self-adjoint extension). The crucial thing is, even if the extension is parameterized by some parameters, Friedrichs extension chooses some specific value at the outset when the concerned symmetric operator is defined, hence, this choice is the most natural one. Having able to do Friedrichs extension means we are saved from introducing an unknown parameter and worrying about what specific value should it be given while doing the extension, which is indeed a problem for anisotropic models \cite{sridip1,sridip2}. It deserves mention that this non uniqueness of self-adjoint extension is pretty common in Physics. For example, the non-uniqueness appears while solving wave equation on a string and we have a family of self-adjoint extensions characterised by the boundary condition. The familiar Dirichlet and Neumann boundary condition are just two members of that family. Hence, non uniqueness of self-adjoint extension does not make a model inconsistent or ill-defined. Just like the physical scenario (i.e how the wave gets reflected at boundary) determines the boundary condition while solving wave equation on string , here the boundary condition on wavefunction must come as a physical input of the concerned quantum cosmological model. \\

To summarise,
\begin{enumerate}
\item A bunch of operator ordering in different anisotropic models hints at the fact that self-adjoint extension is not unique for anisotropic models. Nonetheless the extension exists and thus resolves the problem of nonunitarity.
\item There exist operator orderings, which lead to a $U(1)$ group of self-ajoint extension, even in isotropic models. Though we have a family of self-adjoint extensions, we can pick up a particularly good one, i.e.,  {\it Friedrich's extension} , which is the only one, having a domain contained in the closure of a quadratic form associated with the Hamiltonian . Moreover, it has the same lower bound as the original operator. Friedrichs extension in fact preserves the ground state energy unlike the other extensions. For example, self-adjoint extension of Bianchi-I model in NS ordering sets the ground state energy to a particular finite value where as the original Hamiltonian before extension does not even admit a ground state with finite energy \cite{sridip4}.
\item There is no apparent notion of  {\it Friedrichs extension} in anisotropic models, since the $H_{g}$ for anisotropic models is a not even a semi-bounded operator.
\end{enumerate}

Hence, we conclude that not only the alleged pathology of non-unitary evolution for anisotropic quantum cosmology is actually incorrect, but also the non-uniqueness of self-adjoint extension has nothing to do with anisotropic model, it can happen even in isotropic model. Thus we have successfully established the fact that the quantization scheme does not have inherent inconsistency (concerning non-unitarity) while applied to anisotropic cosmological models. It will be interesting to explore the implications of non-existence of  {\it Friedrichs extension} in anisotropic models in detail, with an aim to find out some \textit{natural} extension, if there is any. \\

It deserves mention that we have assumed spatial homogeneity in our work. It would be worthwhile to extend the analysis to non-homogeneous models where all the operators become fields, function of space-time. The other direction of generalization will certainly be to include rotations in the space time where the implementation of Schutz  formalism will involve two more thermodynamic potentials which are safely ignored for a spacetime without rotation\cite{rubakov}. \\

\appendix
\section{Friedrichs' Extension Theorem}
Friedrichs' extension theorem states:
{\it
Let $A$ be a positive symmetric operator  and let $q\left( \phi ,\psi\right) = \langle \phi , A\psi\rangle$ for $\phi,\psi \in D(A)$. Then $q$ is a closable quadratic form and its closure $\tilde{q}$ is the quadratic form of a unique self-adjoint operator $\tilde{A}$, such that lower bound of its spectrum is lower bound of $q$ and $\tilde{A}$ is the only self-adjoint extension of $A$, whose domain is contained in the form domain of $\tilde{q}$.}\\

We shall go through some defintions and describe how the theorem works. \\

\begin{enumerate}
\item A {\it quadratic form} $q$ is a map $q:\mathcal{Q}(q)\times \mathcal{Q}(q) \rightarrow \mathcal{C}$, where $\mathcal{Q}(q)$ is a dense linear subset of Hilbert space $\mathcal{H}$, which we call {\it Form Domain} such that the map is linear in its second argument and anti-linear in first argument. Basically $q$ takes two arguments from Hilbert space and produce a complex number. 
\item A positive quadratic form $q$ is defined to be one for which $q\left(\psi,\psi\right) \geq 0$ for $\psi\in \mathcal{Q}(q)$.
\item A positive quadratic form is said to be close if $\mathcal{Q}(q)$ is complete under the norm $||\psi||_{form\ norm}^{2} \equiv ||\psi||^{2} +  q\left(\psi ,\psi\right)$, called as {\it Form Norm}.
\end{enumerate}

If $A$ is symmetric and positive, we can define a positive symmetric quadratic form $q$, defined by
\begin{equation}
 q\left( \phi ,\psi\right) = \langle \phi , A\psi\rangle
\end{equation}
with $\mathcal{Q}(q)=\mathcal{D}(A)$, $\mathcal{D}(A)\subset\mathcal{H}$ being the domain of operator $A$.\\

Under the Form norm, we can construct Cauchy sequences in $\mathcal{D}(A)$, and consider the set containing all the limit points of every possible Cauchy sequence. Thereby, we complete $\mathcal{D}(A)$ under the Form-norm to obtain a new Hilbert space $\mathcal{H}_{form}$. Subsequently, $q$ extends to a closed form $\tilde{q}$ on  $\mathcal{H}_{form}$. One can show that  $\mathcal{H}_{form}\subset\mathcal{H}$, which completes the construction of $\tilde{q}$ as a closed form on $\mathcal{H}$. The proof of containment is technically involved and we refer to {\it Theorem X.23} \cite{reed} for details. \\

The quadratic form $\tilde{q}$ is uniquely associated with a self-adjoint operator $\tilde{A}$ such that $\mathcal{D}(\tilde{A})\subset \mathcal{Q}(\tilde{q})$.  We refer to {\it Theroem VIII.15} \cite{reed2} for a detailed and complete proof of the existence of such $\tilde{A}$.\\

The boundness of the spectrum can be proved in following way:\\

Given a $\psi \in \mathcal{D}(\tilde{A})\subset\mathcal{Q}(\tilde{q})\subset\mathcal{H}$ , we can exploit denseness of $\mathcal{Q}(q)$ to have a sequence $\left\{\psi_{n}\in \mathcal{D}(A)\right\}$ such that $\psi_{n}\rightarrow\psi$ under normal Hilber space norm and then
\begin{eqnarray}
\nonumber
\langle\psi, \tilde{A}\psi\rangle =\tilde{q}\left(\psi,\psi\right)=\lim_{n\rightarrow\infty} \tilde{q}\left(\psi_{n},\psi_{n}\right)=\lim_{n\rightarrow\infty} q\left(\psi_{n},\psi_{n}\right)
\end{eqnarray}
Since each of the $q\left(\psi_{n},\psi_{n}\right)$ is non-negative, so is the limit, which proves the lower bound of spectrum of $\tilde{A}$ is lower bound of $q$.\\

In the context of isotropic homogeneous quantum cosmological models, the discussion above applies since $A=-\mathcal{H}_{g\ isotropic}$ is a positive symmetric operator. It is worthwhile to mention 
\begin{enumerate}
\item Any self-adjoint extension of symmetric semi bounded operator ($\mathcal{H}_{g\ isotropic}$ is an example of such operator) with finite deficiency indices is bounded below. 
\item If Friedrichs' extension of a positive symmetric operator ($-\mathcal{H}_{g\ isotropic}$ is an example of such operator) happens to be the only one that is bounded below, then the concerned operator is essentially self-adjoint i.e deficiency indices are $0$.
\end{enumerate}


\begin{thebibliography}{99}
\bibitem{nelson1}N. Pinto-Neto and J.C. Fabris, Class. Quant. Grav. {\bf 30}, 143001 (2013).
\bibitem{wilt} D. L. Wiltshire, in {\it D. L. Wiltshire ``Canberra 1995, Cosmology"} p. 473; arXiv:gr-qc/0101003
\bibitem{halli} J. J. Halliwell, in {\it Quantum Cosmology and Baby Universes}, edited by S. Coleman, J.B. Hartle, T. Piran and S. Weinberg (World Scientific, Singapore, 1991).
\bibitem{lidsey} J.E. Lidsey, Phys. Lett B{\bf 352}, 207 (1995).
\bibitem{nelson2}  N. Pinto-Neto, A. F. Velasco and R. Collistete Jr, Phys. Lett A{\bf 277}, 194 (2000).
\bibitem{kuchar1} K.V. Kuchar, in {\it Conceptual problems in quantum gravity}, edited by A. Ashtekar and J. Stachel (Birkhause, Boston, 1991).
\bibitem{isham} C.J. Isham in {\it Integrable Systems, Quantum Groups and Quantum Field Theory}, edited by L.A. Ibort, M.A. Rodriguez (Kluwer, Dordrecht, 1993).
\bibitem{rovelli} C. Rovelli, Found Phys. {\bf 41}, 1475 (2011), arXiv:gr-qc/0903.3832.
\bibitem{anderson} E. Anderson, in {\it Classical and Quantum Gravity: Theory, Analysis and Applications,} edited by V. R. Frignanni (Nova, New York, 2012), Chap 4; arXiv:gr-qc/1009.2157.
\bibitem{schutz1} B.F. Schutz, Phys. Rev. D {\bf 2}, 2762 (1970).
\bibitem{schutz2} B.F. Schutz, Phys. Rev. D {\bf 4}, 3559 (1971).
\bibitem{rubakov}  V.G. Lapchinskii and V.A. Rubakov, Theor. Math. Phys. {\bf 33}, 1076 (1977).
\bibitem{alvarenga3} F.G. Alvarenga, A.B. Batista, J.C. Fabris, N.A. Lemos and S.V. B. Goncales, Gen. Relativ. Gravit. {\bf 35}, 1639 (2003).
\bibitem{barun}B. Majumder and N. Banerjee, Gen. Relativ. Gravit. {\bf 45}, 1 (2013).
\bibitem{sridip1}  S. Pal and N. Banerjee, Phys. Rev. D {\bf 90}, 104001 (2014).
\bibitem{sridip2} S. Pal and N. Banerjee, arXiv[gr-qc]: 1411.1167, Phys. Rev. D. {\bf 91} 044042 (2015).
\bibitem{sridip3} S. Pal and N. Banerjee, arXiv:1506.02770, Class. Quant. Grav. {\bf 32}, 205005 (2015).
\bibitem{sridip4} S. Pal, arXiv:1504.02912, accepyed in Class. Quant. Grav. , in print
\bibitem{almeida} C.R. Almeida, A.B. Batista, J.C. Fabris and P.R.L.V. Moniz, arXiv:1501.04170.
\bibitem{bastos} C. Bastos, O. Bertolami, N.C. Dias and J.N. Prata, Phys. Rev. D, {\bf 78}, 023516 (2008).
\bibitem{reed}  M. Reed and B. Simon, {\it Methods of Modern Mathematical Physics}, 2nd Edition, Volume 2, (Academic Press, INC. 1975).
\bibitem{reed2}  M. Reed and B. Simon, {\it Methods of Modern Mathematical Physics}, 2nd Edition, Volume 1, (Academic Press, INC. 1975)
\end{thebibliography}
\end{document}